\documentclass[aps,preprint,nofootinbib]{revtex4-1}

\usepackage[utf8]{inputenc}
\usepackage{amsfonts}
\usepackage{color}

\usepackage{amssymb}
\usepackage{amsmath}
\usepackage{stmaryrd}
\usepackage{latexsym}

\usepackage{xcolor}
\definecolor{myurlcolor}{HTML}{08457E}
\definecolor{mylinkcolor}{HTML}{2A52BE}
\definecolor{mycitecolor}{HTML}{E30022}

\usepackage{float}
\usepackage[colorlinks, linkcolor=mylinkcolor, citecolor=mycitecolor, urlcolor=myurlcolor, linktocpage=true]{hyperref}

\def\equationautorefname~#1\null{(#1)\null}
\def\tableautorefname~#1\null{(#1)\null}
\def\figureautorefname~#1\null{(#1)\null}
\def\sectionautorefname~#1\null{(#1)\null}

\let\origref\autoref
\def\autoref#1{\textbf{\origref{#1}}}

\let\origcite\cite
\def\cite#1{\textbf{\origcite{#1}}}

\usepackage{multirow}

\usepackage{tikz}

\usepackage{titlesec}
\titleformat*{\section}{\centering\small\bfseries\scshape}
\titleformat*{\subsection}{\small\bfseries\scshape}
\titleformat*{\subsubsection}{\small\bfseries\scshape}

\newcommand{\be}{\begin{equation}}
\newcommand{\ee}{\end{equation}}
\newcommand{\bea}{\begin{eqnarray}}
\newcommand{\eea}{\end{eqnarray}}
\newcommand{\benn}{\begin{eqnarray*}}
\newcommand{\eenn}{\end{eqnarray*}}

\def\bse{\begin{subequations}}%
\def\ese{\end{subequations}}%

\def\L{{\cal L}}      
\def\S{{\cal S}}  
\def\R{{\cal R}}     

\newcommand{\sfrac}[2]{\dfrac{\,#1\,}{\,#2\,}}

\newcommand{\der}[2]{\sfrac{d #1}{d #2}}

\let\oldsqrt\sqrt
\def\sqrt{\mathpalette\DHLhksqrt}
\def\DHLhksqrt#1#2{%
	\setbox0=\hbox{$#1\oldsqrt{#2\,}$}\dimen0=\ht0
	\advance\dimen0-0.4\ht0
	\setbox2=\hbox{\vrule height\ht0 depth -\dimen0}%
	{\box0\lower0.4pt\box2}}

\def\apj{Astrophysical Journal}

\def\prd{Physical Review D}
\def\prc{Physical Review C}

\def\aap{Astronomy and Astrophysics}

\begin{document}

\title{Neutron Star Structure in the Presence of Nonminimally Coupled Scalar Fields}

\author{A. Sava{\c s} Arapo{\u g}lu}
\email{arapoglu@itu.edu.tr}

\author{K. Yavuz Ek{\c s}i}
\email{eksi@itu.edu.tr}

\author{A. Emrah Y{\"u}kselci}
\email{yukselcia@itu.edu.tr}
\affiliation{Istanbul Technical University, Department of Physics, 34469 Maslak, Istanbul, Turkey \vspace{2cm}}

\begin{abstract}
We study the structure of neutron stars in scalar-tensor theories for the nonminimal coupling of the form $(1+\kappa \xi \phi^{2})\R$. We solve the hydrostatic equilibrium equations for two different types of scalar field potentials and three different equations of state representative of different degrees of stiffness. We obtain the mass-radius relations of the configurations and determine the allowed ranges for the term $\xi\phi^2$ at the center of the star and spatial infinity based on the measured maximum value of the mass for neutron stars and the recent constraints on the radius coming from gravitational wave observations. Thus we manage to limit the deviation of the model from general relativity. We examine the possible constraints on the parameters of the model and compare the obtained restrictions with the ones inferred from other cosmological probes that give the allowed ranges for the coupling constant only. In the case of the Higgs-like potential, we also find that  the central value for the scalar field cannot be chosen arbitrarily but it depends on the vacuum expectation value of the field.  Finally, we discuss the effect of the scalar field potential on the mass and the radius of the star by comparing the results obtained for the cases considered here.
\end{abstract}

\maketitle

\raggedbottom

\section{INTRODUCTION}
Neutron stars constitute an important class of extreme relativistic objects allowing one to examine the theories of gravitation in the strong field regime \cite{psaltis-rw-2008,Stairs:2003}. Their supranuclear densities stemming from their large masses and small radii make them also interesting laboratories for matter under extreme conditions, which is modelled by different equations of state (EOS) \cite{EoS1,EoS2,EoS3}. Such extreme conditions make the determination of EOS a difficult problem in nuclear physics but, on the other hand, the studies in this direction may allow one, for example, to constrain even dark matter models in a way that is inaccessible to terrestrial or cosmological probes \cite{prl-121-061801,prl-121-061802}. The stability and the structure of neutron stars depend both on their possible EOS and the relativistic theory of gravity being used in their description. Although mostly Einstein's general relativity (GR) is the theory used in this setting, the recent observations \cite{perlmutter1998,riess1998,planck2015} lead to considering alternative theories of gravity. The greater mass-radius ratio of neutron stars which is basically characterized by the compactness parameter $GM/(Rc^{2})$ and curvature, implies that their gravitational fields are many orders of magnitude stronger, for example, than that of the Sun and so are in a regime that cannot be probed by the solar system tests and cosmological observations \cite{psaltis-rw-2008,spectra_psaltis,degeneracy1}.

Scalar-tensor theories are one of the most natural extensions of GR \cite{fierz1956,jordan1959,brans-dicke1961,bergmann1968,nordtvedt1970,wagoner1970}. Different models in this sort of extension of GR are classified by the type of the nonminimal coupling of scalar field to Ricci scalar and the type of the scalar field potential in the action. They arise naturally in the low energy limit of fundamental theories like string theory and other higher dimensional theories upon compactification. Such theories are used extensively in cosmology for describing both the inflationary epoch \cite{faraonibook,Buck2010,Bezrukov2008,Calmet2018} and the current accelerated expansion of the Universe \cite{sadatian2013}.

Including a nonminimal coupling term in the action in some cases is not just a matter of choice but is dictated by the quantum corrections to the scalar field in curved spacetime \cite{chernikov1968,callan-etal1970,birrell-davies1980,birrell-davies-1982}. In inflationary models, inclusion of a nonminimal coupling of the form $\xi \phi^{2} \R$ in the matter sector, where $\xi$ is a dimensionless coupling constant, is expected on fairly general grounds. For instance, in successful reheating models the inflaton couplings, even if they are weak, can induce a nontrivial $\xi$; that is, even if one starts with $\xi=0$, a nonminimal coupling of the aforementioned form will be radiatively generated. Such a type of nonminimal coupling is studied in the cosmological setting to provide viable models that are ruled out in the case of minimal coupling corresponding to the case $\xi=0$ \cite{Komatsu1998,Komatsu1999,Tsujikawa2000,fujii_maeda_2003,Boubekeur2015}. A common point in all these studies is that the dynamics is studied in the Einstein frame which is obtained by applying conformal transformations to the original system in the Jordan frame and putting limits on the value of $\xi$ to make the models compatible with the observations. Here we work directly in the Jordan frame without discussing the equivalence of the Jordan and Einstein frame formulations of the theory. Previously, this approach has been followed in the context of neutron stars with the scalar potential $V(\phi)\!=\!0$ in Ref.\ \cite{salgado1998} for the phenomena of spontaneous scalarization and in Ref.\ \cite{kazanas2014} only for $\xi\!=\!1/6$ corresponding to the conformal coupling case, which is the only value of the coupling constant allowed by the equivalence principle. Besides the zero potential case in Ref.\ \cite{fuzfa1} a Higgs-like potential has been considered in the same model for a static and spherically symmetric configuration with a constant density. The same approach has been also followed in Ref.\ \cite{hrycyna_what_xi} as an application of the dynamical system analysis to a simple cosmological model with a constant potential function.

Structure of neutron stars in the presence of scalar-tensor fields had been investigated in Refs.\ \cite{zag92,har98,hor11,kazanas2014,sot18}, including the rotating case \cite{don+13,don16,staykov2016,don+18,sta+18}. Damour and Esposito-Far\`{e}se \cite{Damour:1992,damour1993,damour_ppn,farese2004} discovered the effect of spontaneous scalarization beyond a critical compactness, which, more recently, was reconsidered by the authors of Refs.\ \cite{alt+17,mor17}. Spontaneous scalarization as a Higgs-like mechanism in which a photon gains mass is studied in Ref.\ \cite{fra+18} and with nonminimal coupling in Ref.\ \cite{coa+17}. Minamitsuji and Silva \cite{min16} examined the effect of disformal coupling and Cisterna \textit{et al}.\ \cite{cis+15} studied nonminimal derivative coupling. Further applications such as pulse profiles of pulsars \cite{sot17b,sot18b,yunes-silva2018}, oscillation modes \cite{sot04,sot05,sil+14,yaz+17,men18,alt+18,bla+18,Silva2019}, and oscillations of circumstellar disks \cite{osc_Doneva} are also topics considered actively in the literature.

The aim of the present work is to study the structure of neutron stars in the presence of a scalar field that couples to gravity nonminimally through  a term of the form $\xi \phi^{2} \R$. We first consider the case $V(\phi)\!=\!0$ and put limits on the values of the parameters of the model, specifically, the coupling constant $\xi$ and the central value of the scalar field $\phi_{\rm c}$, choices that allow one to construct a stable relativistic star configuration. We also study the potential of the form $V(\phi)\!=\!\lambda(\phi^{2}-\nu^2)^2/4$ that is mostly considered in cosmological models describing the early and the late time evolutions of the Universe. We derive the field equations and solve the related hydrostatic equilibrium equations for three different EOS, namely, SLy \cite{eos_sly}, WFF1 \cite{eos_wff1} and MS1 \cite{eos_ms1}, and plot the mass-radius ($M(R)$) relations for each case. By considering the maximum observational mass of the neutron stars \cite{J0348} and radius constraints belonging to the observation of \textit{GW170817} \cite{GW170817_observe,LIGO_GW170817}, we determine an allowed range for the term $\xi\phi_{\rm c}^2$, which practically represents the deviation of the model from GR. In order to find restrictions for the parameters individually, the obtained constraints are compared with the ones that come from different cosmological probes. Consequently, we find that the central value of the scalar field cannot be chosen arbitrarily but it strictly depends on the choice of the coupling constant and vice versa. We also review the parametrized post-Newtonian (PPN) formalism \cite{ppn_gamma,ppn_beta_1,ppn_beta_2,ppn_review} and dipole radiation in pulsar-white-dwarf binary systems \cite{dipole_rad_Freire:2012,dipole_rad_Shao:2017} and discuss the emerging constraints briefly.

The caveat in these type of tests of gravity theories is that there is a strong degeneracy between the theory of gravity being used and the EOS; that is, a gravity theory that is ruled out for an EOS may be allowed for another \cite{degeneracy1,degeneracy2,degeneracy3,dgnrcy4_Doneva:2017,dgnrcy5_Shao:2019}. Thus the present work investigates the possible values of $\xi$ in this sense and compares them with other observations.  A possible way of breaking this interplay between the theory of gravity and the EOS is to examine, for example, some electromagnetic effects near the star \cite{psaltis-rw-2008,burst_psaltis,burst_osc_silva,osc_DeDeo,osc_Doneva,osc_Pappas} by considering only the exterior region \cite{yunes-silva2018}.

The plan of the paper is as follows : In Sec.\ \autoref{setup} we derive the field equations and applying the static and spherically symmetric metric we obtain a set of differential equations describing the hydrostatic equilibrium of the system. In Sec.\ \autoref{numerical} we explain the numerical procedure to solve this set of equations and compare our results with different approaches in the literature. Then the results for zero and Higgs-like potentials are examined in Sec.\ \autoref{models} and concluding remarks are given in Sec.\ \autoref{conclusion}.

\section{EQUATIONS OF HYDROSTATIC EQUILIBRIUM} \label{setup}
Action for the nonminimally coupled scalar field in the Jordan frame is 
\begin{equation}
	\S = \int \! d^4 x \, \sqrt{|g|} \, \bigg[ \sfrac{1}{\, 2 \kappa \,} \R + \sfrac{\,1\,}{2} \xi \phi^2 \R - \sfrac{\,1\,}{2} \nabla^c \phi \, \nabla_{\!\!c} \, \phi - V\!(\phi) + \L^m \bigg]
\label{eq:action}
\end{equation}
where $\kappa = 8\pi$ and $\xi$ is the coupling constant.\footnote{We use geometrical units ($G = c = 1$) throughout this study.} Variation of the action with respect to the metric can be written in the form
\begin{equation}
	 \R_{\mu \nu} - \sfrac{\,1\,}{2} \R \, g_{\mu \nu} = \, \kappa_{\rm eff} \big[ T_{\mu\nu}^{(m)} + T_{\mu\nu}^{(\phi)} \big]
\label{eq:fieldeq1}
\end{equation}
where $\kappa_{\rm eff}$, energy-momentum tensors for the fluid $T_{\mu\nu}^{(m)}$ and for the scalar field $T_{\mu\nu}^{(\phi)}$ are
\begin{equation}
	\kappa_{\rm eff}(\phi) = \kappa \big( 1 + \kappa \xi \phi^2 \big)^{-1} \,\, ,
\label{eq:kappa_eff}
\end{equation}
\begin{equation}
	T_{\mu\nu}^{(m)} = \big(\rho + P\big) u_\mu u_\nu + P g_{\mu\nu} \,\, ,
\label{eq:matter_tensor}
\end{equation}
\begin{equation}
	T_{\mu\nu}^{(\phi)} = \nabla_{\!\!\mu} \, \phi \, \nabla_{\!\nu} \, \phi \\[2pt] - g_{\mu \nu} \bigg[ \sfrac{1}{2} \, \nabla^c \phi \, \nabla_{\!\!c} \, \phi + \, V\!(\phi) \bigg] - \xi \big( g_{\mu \nu} \boxempty - \nabla_{\!\!\mu} \, \nabla_{\!\nu} \big) \phi^2 \: .
\label{eq:scalar_tensor}
\end{equation}
It is possible to interpret the field equations for the nonminimally coupled scalar field in different conventions. In this paper we follow the one called \textit{the effective coupling approach} referring to the fact that the expression given in Eq.\ \autoref{eq:kappa_eff} represents an effective gravitational coupling for the total energy-momentum tensor (for a detailed discussion see Ref.\ \cite{faraonibook}). In this case, the energy-momentum tensor of the fluid is conserved in the sense that $\nabla^{\mu} T_{\mu\nu}^{(m)} = 0$.

The equation of motion for the scalar field is obtained as
\begin{equation}
	\boxempty \! \phi + \xi \phi \R - \dfrac{dV\!(\phi)}{d \phi} = 0 \, .
\label{eq:fieldeq2}
\end{equation}

We use static and spherically symmetric metric 
\begin{equation}
	ds^2 = - e^{2f(r)} dt^2 + e^{2g(r)} dr^2 + r^2 \big(d\theta^2 + \sin^2\theta \, d\varphi^2\big),
\label{eq:metric}
\end{equation}
where $f(r)$ and $g(r)$ are the functions that will be determined. Using this metric, we get the following set of equations
\begin{subequations}
	\begin{align}
	g' &= \bigg[ \sfrac{1}{r} \!+\! \kappa_{\rm eff} \xi \phi \phi' \bigg]^{-1} \bigg[ \sfrac{1 \!-\! e^{2g}}{2r^2} \!+\! \kappa_{\rm eff} \bigg\{ \sfrac{1}{2} \big(\rho \!+\! V\big)e^{2g} + \xi \phi \bigg( \phi'' \!+\! \sfrac{2\phi'}{r} \bigg) \!+\! \bigg(\xi \!+\! \sfrac{1}{4}\bigg)\phi'^2 \bigg\} \bigg] \label{eq:components_tt} \: , \\[3mm] 
	f' &= \bigg[ \sfrac{1}{r} \!+\! \kappa_{\rm eff} \xi \phi \phi' \bigg]^{-1} \bigg[ \!-\! \sfrac{1 \!-\! e^{2g}}{2r^2} + \kappa_{\rm eff} \bigg\{ \sfrac{1}{2} \big(P \!-\! V\big)e^{2g} + \sfrac{1}{4}\phi'^2 - \sfrac{2\xi\phi\phi'}{r} \bigg\} \bigg] \label{eq:components_rr} \: , \\[3mm] 
	f'' &= -\big(f'\!-\!g'\big) \Big[ f' \!+\! \sfrac{1}{r} \Big] \!+\! \kappa_{\rm eff} \bigg\{ \! \big(P\!-\!V\big)e^{2g} \!-\! 2\xi\phi \bigg[ \phi'' \!+\! \phi' \bigg( \! f'\!-\!g'\!+\!\sfrac{1}{r} \! \bigg) \! \bigg] \!-\! \bigg( \! 2\xi \!+\! \sfrac{1}{2}\bigg) \phi'^2 \bigg\} \: , \label{eq:components_QQ}
	\end{align}
\label{eq:components_eq}%
\end{subequations}
which are obtained from $tt$, $rr$, and $\theta\theta$ components of Eq.\ \autoref{eq:fieldeq1}, respectively. Here, prime denotes the derivative with respect to coordinate radius $r$.  The equation of motion for the scalar field given in Eq.\ \autoref{eq:fieldeq2} becomes
\begin{equation}
	\phi'' + \bigg[f' \!-\! g' \!+\! \sfrac{2}{r}\bigg]\phi' - 2\xi\phi \bigg[f'' \!+\! \big(f'\!-\!g'\big) \Big( f' \!+\! \sfrac{2}{r} \Big) + \sfrac{1 \!-\! e^{2g}}{r^2}\bigg] - e^{2g} \der{V}{\phi} = 0 \,\, ,
\label{eq:scalar_eq}
\end{equation}
and the energy-momentum conservation equation of the fluid ($\nabla^{\mu} T_{\mu\nu}^{(m)} = 0$) yields
\begin{equation}
	P' = - f'\big(P + \rho\big)  \,\, .
\label{eq:pressure_eq}
\end{equation}

In the absence of a scalar field the mass function is defined as
\begin{equation}
	m(r) \equiv \sfrac{r}{2} \big(1 - e^{-2g}\big) \: ,
\end{equation}
and it gives the total mass of the configuration when calculated on the surface of the star ($R$) i.e. $M\!=\!m(R)$.  However, the presence of a scalar field leads to a change since its existence outside the star contributes to the measured mass. In scalar-tensor theories it is possible to consider various mass types for different situations, such as ADM \cite{adm_mass} and Komar \cite{komar_mass} masses, which coincide for static configurations \cite{mass_coincides}. We therefore prefer to use the mass function for the configuration similar to the one given in Ref.\ \cite{salgado1998} in the following form
\begin{equation}
	M \equiv \lim_{r \rightarrow \infty} m(r) = 4\pi \int_{0}^{\infty} r^2 E(r) \, dr
\label{eq:mass}
\end{equation}
where the function $E(r)$ is defined as 
\vspace{2mm}
\begin{equation}
	E(r) \equiv \sfrac{\kappa_{\rm eff}}{\kappa} \bigg[ \rho + \sfrac{1}{2} \big( \phi' e^{-g} \big)^2 + V(\phi) + 2\xi \bigg\{ \phi \bigg( \! \phi'' + \phi' \bigg[ \sfrac{2}{r} \!-\! g' \bigg] \bigg) + (\phi')^2 \bigg\} e^{-2g} \bigg] \: .
\label{eq:mass_E}
\end{equation}
Note that this expression contains energy densities of the fluid and the scalar field together with an additional term depending on the nonminimal coupling. Although the normal matter does not contribute to this expression in the exterior region of the star, the scalar field contributes still to the mass integral. That is why the integration has to be carried out from the center of the star to spatial infinity, which, in practice, should not be necessary since the scalar field has to reach its asymptotic value at a point that is not too far from the surface. Moreover, the gravitational mass defined in Ref.\ \cite{kazanas2014} is consistent with ours.

It is noteworthy to point out that the asymptotic value of the scalar field ($r \! \rightarrow \! \infty$) needs an extra attention when, in particular, the cases with nonzero potentials are investigated. Appropriate central values for the scalar field have to be chosen such that the constant asymptotic value of the scalar field satisfies $V(\phi_\infty) \!=\! 0$. On the other hand, such a restriction for a cosmological background can be modified in a way that the asymptotically nonvanishing potential can be interpreted as an effective cosmological constant \cite{sotiriou}. There is also an additional condition obtained by rearranging Eq.\ \autoref{eq:fieldeq2} as follows
\begin{equation}
	\bigg[1 + \sfrac{6\kappa\xi^2 \phi^2}{1 + \kappa \xi \phi^2}\bigg] \! \boxempty \! \phi = -\sfrac{\kappa \xi \phi}{1 + \kappa \xi \phi^2} \Big[ \rho - 3P + \big( 1+6\xi \big) \nabla^c \phi \, \nabla_{\!\!c} \, \phi + 4V \Big] + \der{V}{\phi} \, ,
	\label{eq:fieldeq3}
\end{equation}
and this implies that, in order to get a stable configuration, the potential of the scalar field has to satisfy the following expression at spatial infinity
\begin{equation}
	\bigg[\! -\sfrac{4 \kappa \xi \phi}{1 + \kappa \xi \phi^2} V(\phi) + \der{V(\phi)}{\phi} \bigg] \bigg|_{\phi = \phi_\infty} = 0 ,
\end{equation}
which is directly related to the last term in Eq.\ \autoref{eq:mass_E}.

\section{THE OBSERVATIONAL CONSTRAINTS AND NUMERICAL PROCEDURE}  \label{numerical}
In order to solve the system numerically, we need to construct a first-order differential equation set that describes the configuration. Plugging $f''$ from Eq.\ \autoref{eq:components_QQ} into Eq.\ \autoref{eq:scalar_eq} and using Eqs.\  \autoref{eq:components_tt}, \autoref{eq:components_rr}, and \autoref{eq:scalar_eq} together with Eq.\ \autoref{eq:pressure_eq} the desired set of equations can be obtained. Furthermore, an EOS, which relates density and pressure of the fluid, has to be provided to have a closed dynamical system. We use three different types, namely SLy \cite{eos_sly}, WFF1 \cite{eos_wff1}, MS1 \cite{eos_ms1}, calculations of which are based on Skyrme models, microscopic calculations, and the relativistic mean field theory, respectively. As these EOS are in the form of tabulated data they are incorporated into the solution by numerical interpolation.

The system described above is solved numerically starting from the center of the star ($r\!=\!0$) up to spatial infinity while the surface of the star is determined by the condition $P(R) \!=\! 0$ during the process. Moreover, we need to specify some initial conditions for the functions to begin with: we set $P(0) \!=\! P_{\rm c}$, $\phi(0) \!=\! \phi_{\rm c}$, $\phi'(0) \!=\! 0$, $f(0) \!=\! f_{\rm c}$, $g(0) \!=\! 0$, and $M(0) \!=\! 0$ requiring regularity conditions at the center. At spatial infinity, on the other hand, we have $g(r_\infty) \!=\! 0$ and $f(r_\infty) \!=\! 0$ so that the metric becomes Minkowskian as $r \! \rightarrow \! \infty$. We note that it is necessary to impose an initial condition different than zero for the metric function $f(r)$ in order to ensure that its asymptotic value matches the required asymptotic type of the metric. 

Another crucial issue is matching the interior and the exterior solutions at the surface of the star. After the solution reaches the surface of the star, we continue to integrate the same system by imposing $P \!=\! \rho \!=\! 0$. We check afterwards whether the above defined boundary conditions are satisfied; that is, the values of the functions at the surface belonging to the interior solution provide initial conditions for the exterior solution that has the correct asymptotic behavior for the system at spatial infinity. The advantage of this approach is that no approximations for the exterior solution are required. On the other hand, it should be emphasized that the resulting $M(R)$ relations characterized by the EOS and the central value of the scalar field are compatible with the observations only if they allow for a maximum mass exceeding $(2.01\pm0.04)\,\rm M_{\odot}$ \cite{J0348} and obey the radius constraint obtained from \textit{GW170817} event which is $R=(11.9\pm1.4)\,\rm km$ \cite{LIGO_GW170817}.

We will see that the asymptotic value of the scalar field is determined by its central value and its compatibility with other observations can be checked via another method.  In this respect, it may be useful to compare the restrictions on the $\phi_{\rm c}$ and $\xi$ here with the ones coming from PPN formalism and dipole radiation in pulsar-white-dwarf binary systems.  Following the methods of Ref.\ \cite{damour_ppn} one can use
\begin{equation}
	\alpha(\phi) = -\sfrac{\sqrt{2\kappa} \, \xi \phi}{\big[1 + \kappa \xi (1+6\xi)\phi^2 \big]^{1/2}} \qquad \mbox{and} \qquad \beta(\phi) = -\sfrac{2\xi(1+\kappa\xi\phi^2)}{\big[1 + \kappa \xi (1+6\xi)\phi^2 \big]^2}
\label{eq:ppn_1}
\end{equation}
which are related to the PPN parameters as 
\begin{equation}
	\gamma_{\mbox{\tiny PPN}}-1 = \sfrac{-2\alpha_\infty^2}{1 + \alpha_\infty^2} \qquad \mbox{and} \qquad \beta_{\mbox{\tiny PPN}}-1 = \sfrac{\beta_\infty \, \alpha_\infty^2}{2(1+\alpha_\infty^2)^2} \,\, 
\label{eq:ppn_2}
\end{equation}
where $\alpha_\infty \equiv \alpha(\phi_\infty)$ and $\beta_\infty \equiv \beta(\phi_\infty)$. Applying the limits on PPN parameters \cite{ppn_gamma,ppn_beta_1,ppn_beta_2,ppn_review} to these expressions one can estimate that $\alpha_\infty^2 \lesssim 10^{-5}$ and $|\beta_\infty| \, \alpha_\infty^2 \lesssim 1.6 \times 10^{-4}$. On the other hand, constraints on the matter-scalar coupling parameter ($\alpha_\infty$) coming from the analysis of dipole radiation in pulsar-white-dwarf binary systems investigated in Refs.\ \cite{dipole_rad_Freire:2012,dipole_rad_Shao:2017} provide a more stringent bound of $|\alpha_\infty| \lesssim 10^{-4}$. Using Eqs.\ \autoref{eq:ppn_1} and \autoref{eq:ppn_2}, we see that it is possible to write $\alpha_\infty^2 \approx 2 \kappa \xi^2 \phi_\infty^2 \!<\! 10^{-8}$, therefore, the term $\xi^2 \phi_\infty^2$ can be restricted with the help of these observational bounds.

There is another approach that is particularly useful for our purposes here. In Ref.\ \cite{hrycyna_what_xi} the same model has been examined in the context of dynamical system analysis to limit the coupling constant in the light of cosmological observations. The authors have found various restrictions for the value of $\xi$ under certain assumptions but we only consider the range $-2.6051 \!<\! \xi \!<\! -0.0633$, which corresponds to the Universe consisting of both baryonic and dark matter together with a positive cosmological constant. Using this interval for $\xi$ in our analysis where we have constrained the combination $\xi \phi^{2}_c$ we get an allowed range for $\phi_{\rm c}$.

\begin{figure*}[!ht]
	\centering
	
	\begin{tabular}{@{}c@{}}
		\includegraphics[width=.5\linewidth]{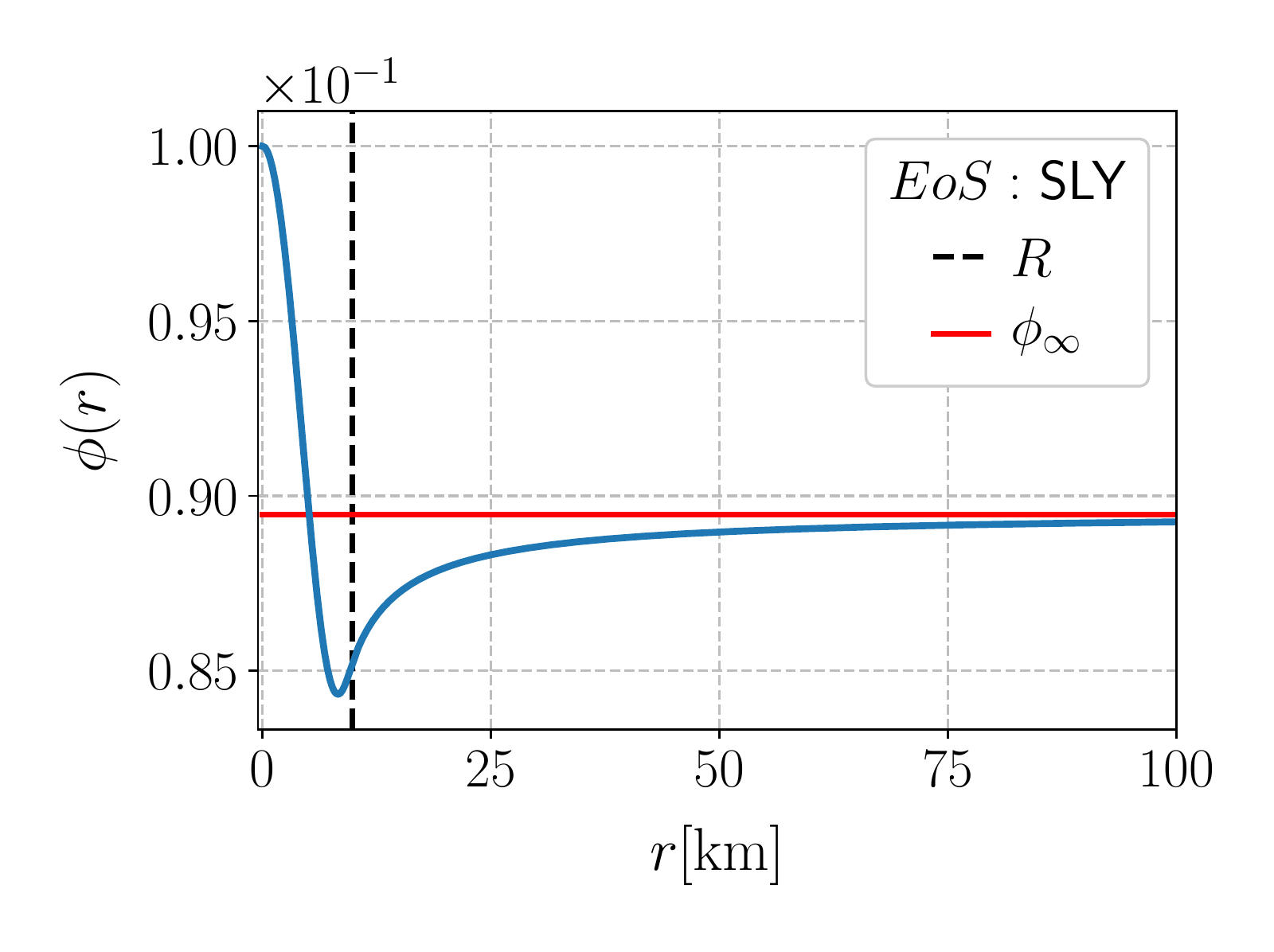} \hspace{-5mm} \\
	\end{tabular}
	\begin{tabular}{@{}c@{}}
		\includegraphics[width=.5\linewidth]{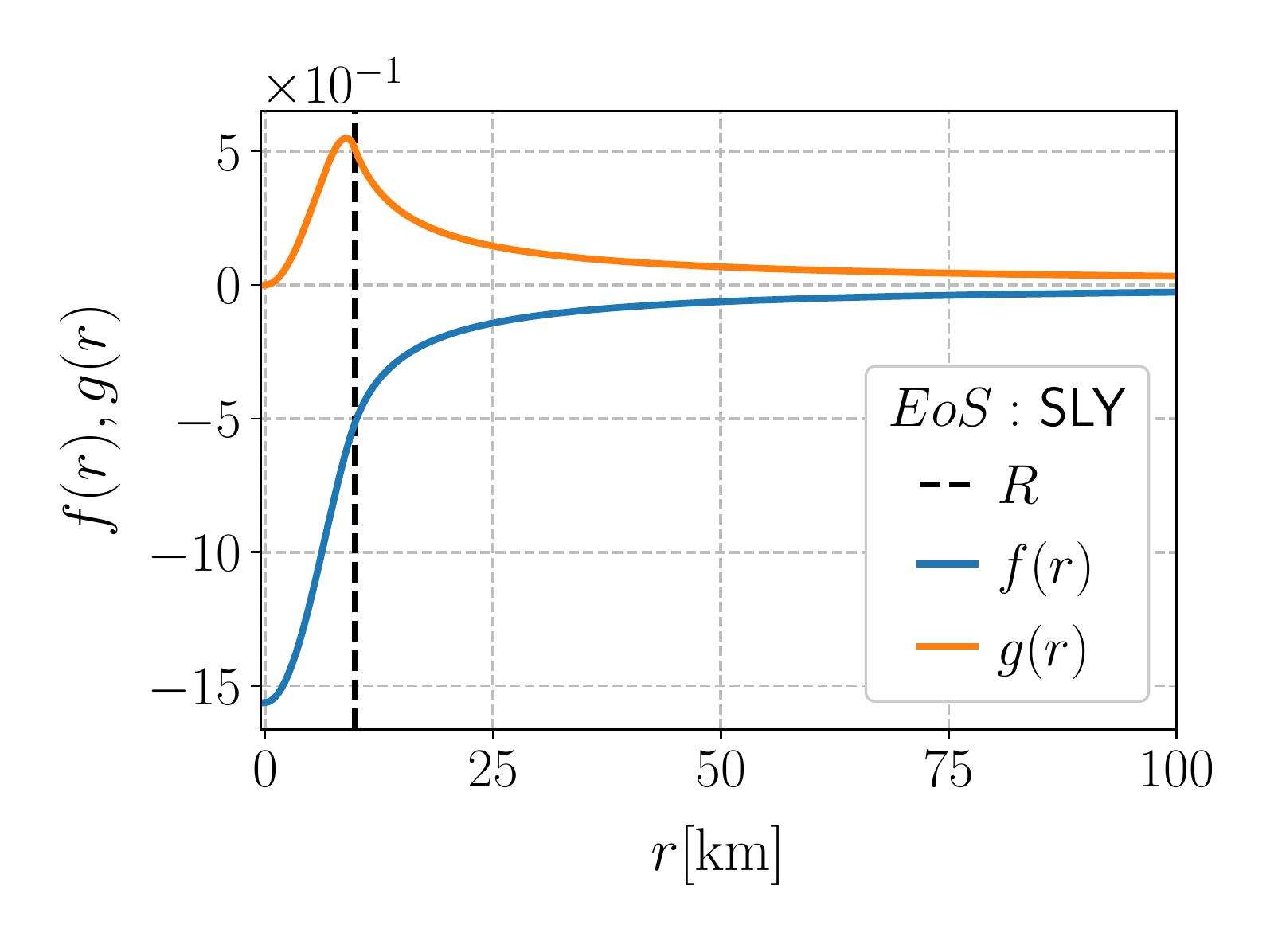} \\
	\end{tabular}
	
	\vspace{-5mm}
	
	\caption{Radial profiles of the scalar field $\phi(r)$ (left) and the metric functions $f(r)$ and $g(r)$ (right) for zero potential. Here $\xi=-0.65$ and $\phi_{\rm c} = 0.1$. $R$ indicates the radius of the star.}
	\label{fig:zero_func}
\end{figure*}

\section{APPLICATIONS OF THE MODEL FOR ZERO AND HIGGS-LIKE POTENTIALS}  \label{models}

\subsection{Zero Potential}
This case was examined in Ref.\ \cite{salgado1998} within the framework of spontaneous scalarization and in Ref.\ \cite{kazanas2014} only for the conformal coupling ($\xi\!=\!1/6$). In Ref.\ \cite{salgado1998} it is reported that only asymptotically vanishing values of the scalar field were taken into account and different terms contributing to the total energy of the configuration were investigated. Since our aim is to discuss the possible constraints on the parameters of the model ($\phi_{\rm c}$ and $\xi$) rather than examining the underlying mechanism, it is not necessary to restrict ourselves with the asymptotic behavior of the scalar field as long as it is compatible with observations. On the other hand, as shown in Ref.\ \cite{kazanas2014}, working with isotropic coordinates allows one to match the numerical interior solution with the analytic exterior solution that is not in one-to-one correspondence with the Schwarzschild coordinates \cite{yunes-silva2018}. Nevertheless, the coherence of the two approaches can be seen easily by comparing the behavior of the metric functions and the scalar field outside the star.

\begin{figure*}[!ht]
	\centering
	\hspace{-2mm}
	\begin{tabular}{@{}c@{}}
		\includegraphics[width=.45\linewidth]{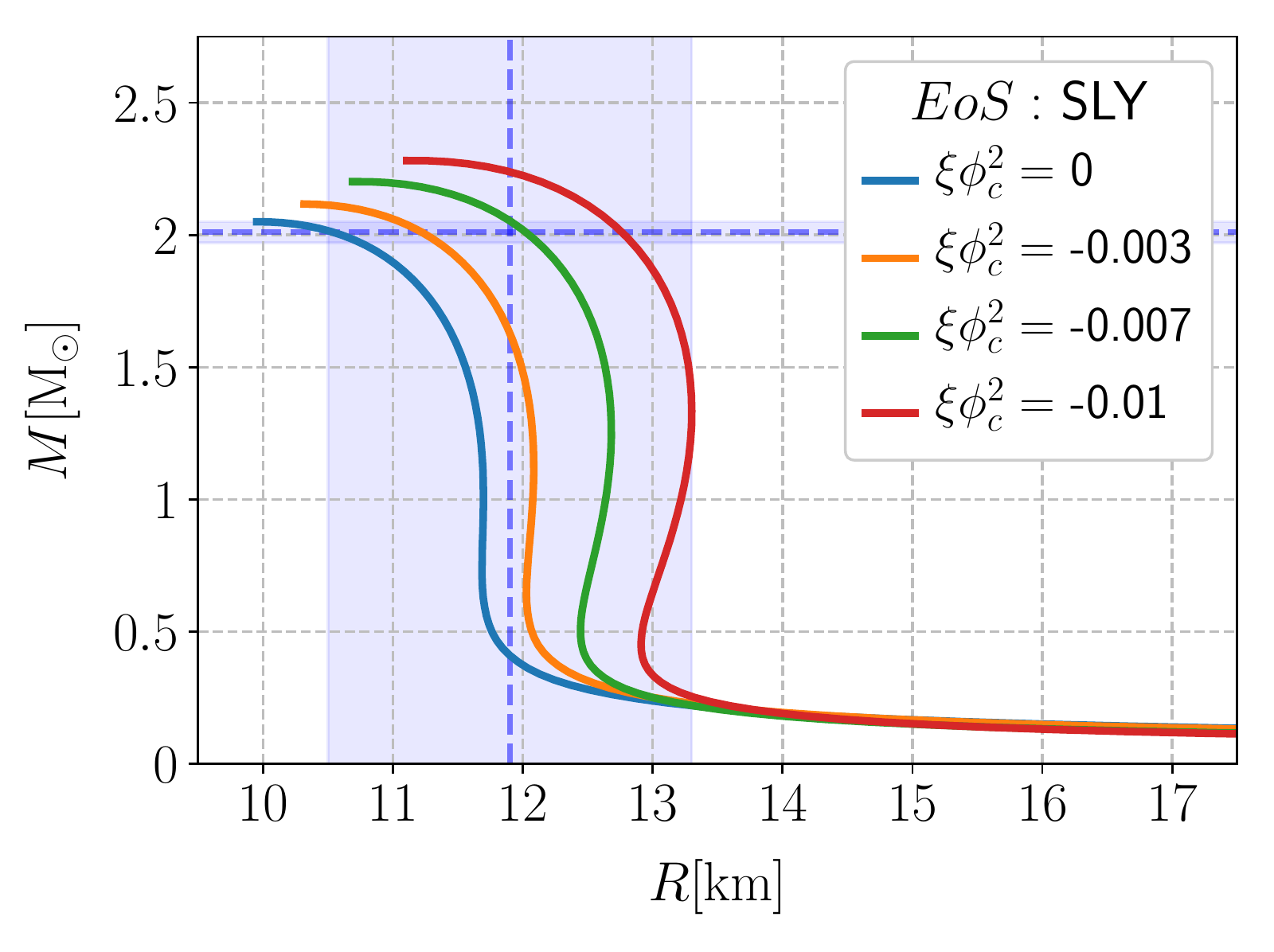} \\
		\label{fig:mr_sly01_V0}
	\end{tabular}
	\begin{tabular}{@{}c@{}}
		\includegraphics[width=.45\linewidth]{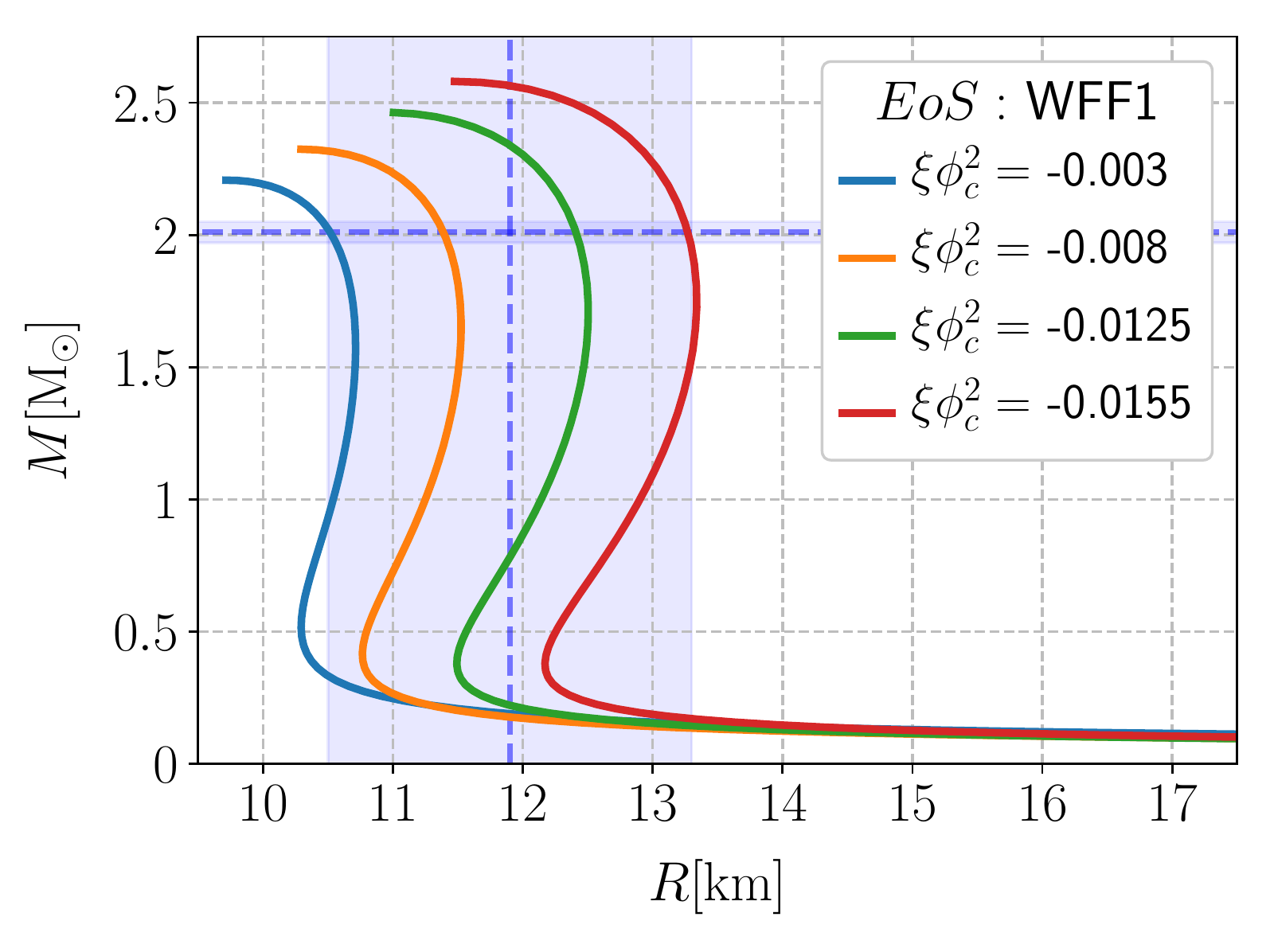} \\
		\label{fig:mr_sly1_V0}
	\end{tabular}
	
	\vspace{-8mm}
	
	\begin{tabular}{@{}c@{}}
		\includegraphics[width=.47\linewidth]{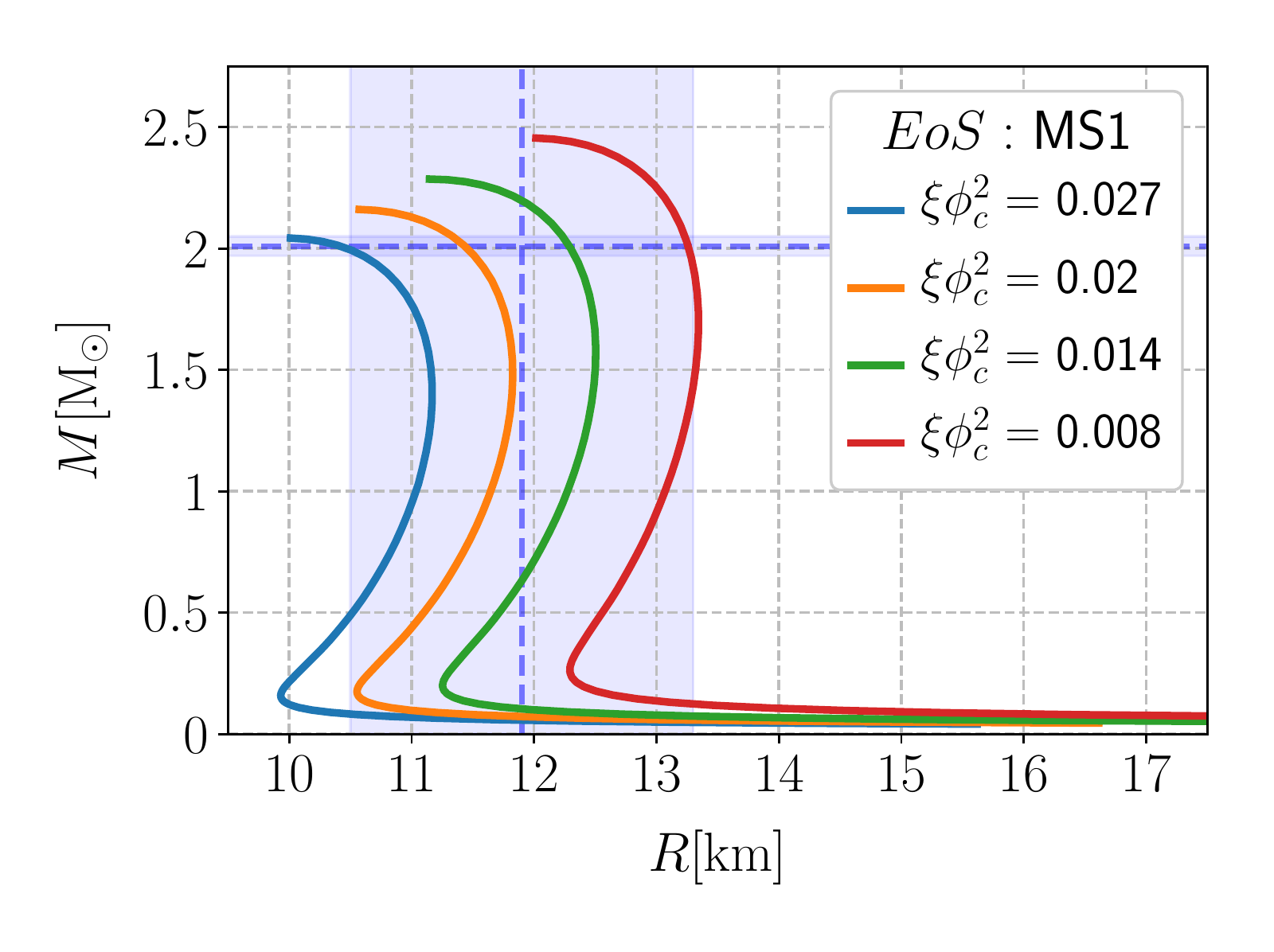} \\
		\label{fig:mr_wff101_V0}
	\end{tabular}
	
	\vspace{-8mm}
	
	\caption{Mass-radius relations of the configuration with zero potential. Here $\phi_{\rm c} \!=\! 0.1$ and $\xi \phi_{\rm c}^2 \!=\! 0$ represents GR solutions. The horizontal dashed line is $(2.01\pm0.04)\,\rm M_{\odot}$ for \textit{J0348+0432} \cite{J0348}, whereas the vertical dashed line, $R \!=\! (11.9 \pm 1.4) \, \rm km$, stands for the radius constraint obtained from the \textit{GW170817} event \cite{LIGO_GW170817}.}
	\label{fig:mr_zero} \vspace{5mm}
\end{figure*}

\begin{figure*}[!ht] 
	\centering
	\begin{tabular}{@{}c@{}}
		\includegraphics[height=.32\linewidth]{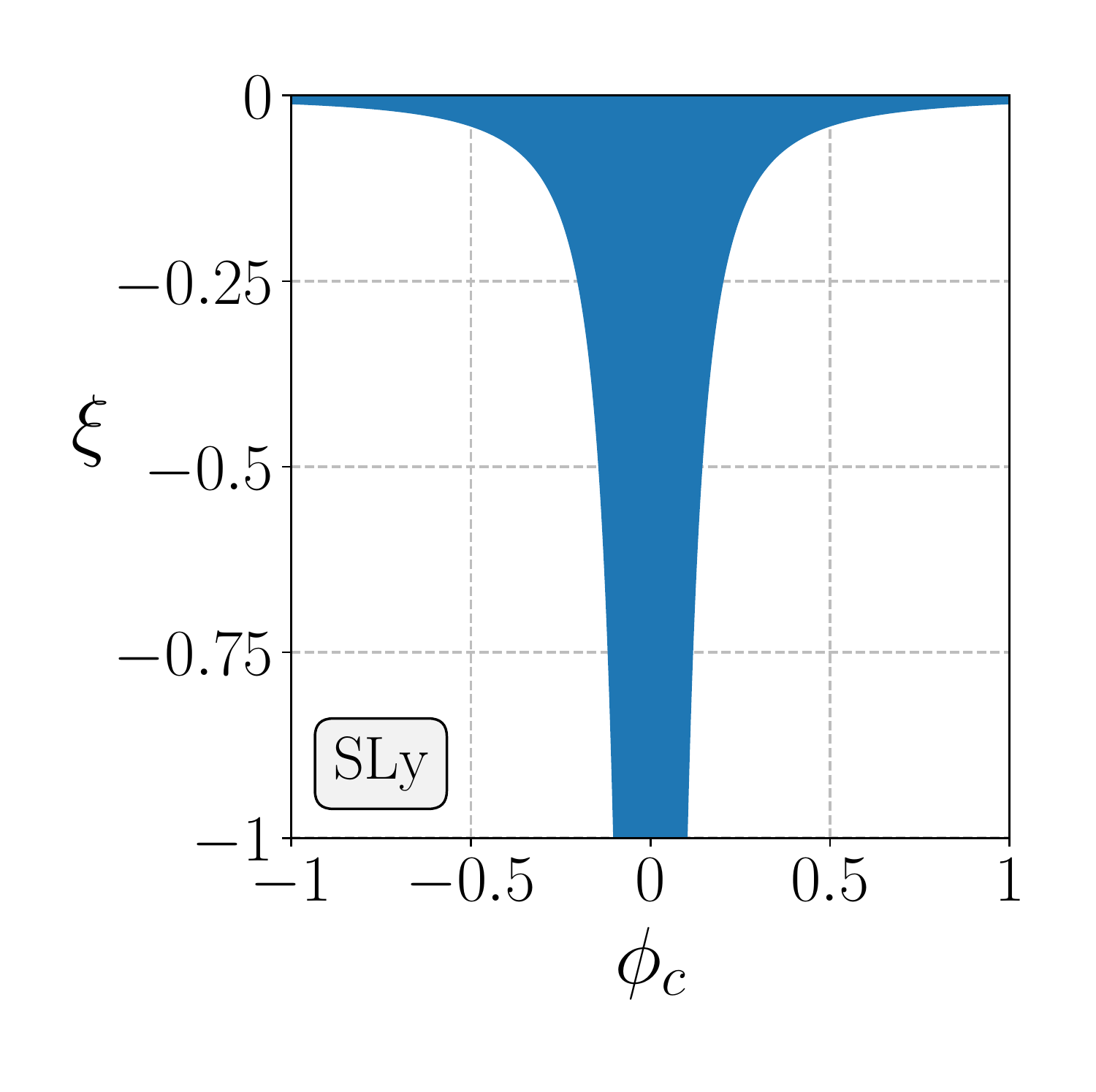}
		\label{fig:fig_SLY_constraint}
	\end{tabular}
	\begin{tabular}{@{}c@{}}
		\includegraphics[height=.32\linewidth]{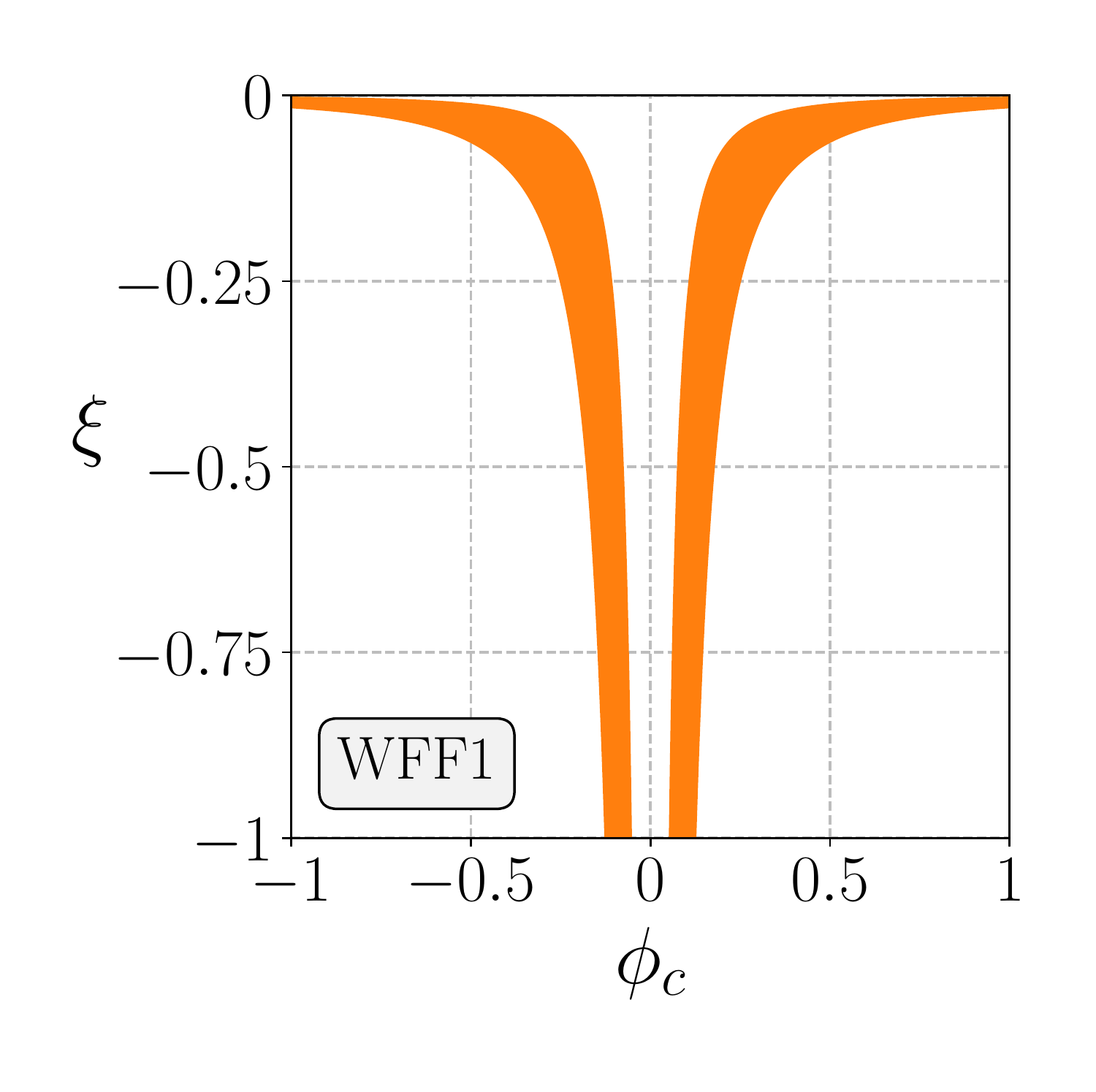}
		\label{fig:fig_WFF1_constraint}
	\end{tabular}
	\begin{tabular}{@{}c@{}}
		\includegraphics[height=.32\linewidth]{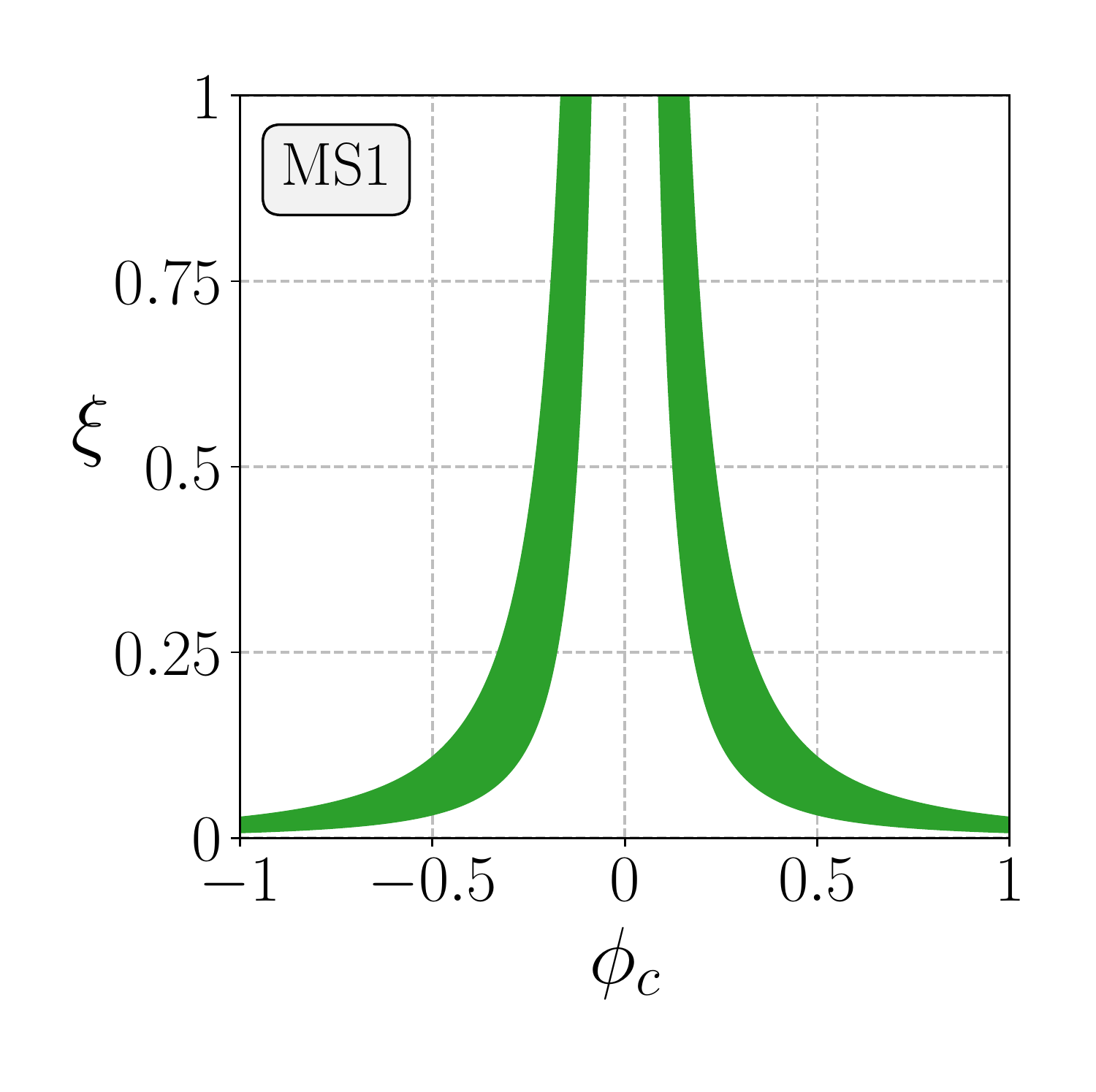}
		\label{fig:fig_MS1_constraint}
	\end{tabular}

	\vspace{-6mm}
	
	\caption{Parameter space of the constraints given in Fig.\ \autoref{fig:mr_zero}. The ranges of the term $\xi\phi^2_{c}$ are ($-0.01, 0$), ($-0.0155, -0.003$), and ($0.008, 0.027$) for SLy, WFF1, and MS1 from left to right, respectively.}
\label{fig:constraints}
\end{figure*}

Fig.\ \autoref{fig:zero_func} shows the characteristic radial profile of the metric functions and the scalar field for a sample EOS. As can be seen from the figure, behavior of the functions at spatial infinity satisfies the asymptotic flatness condition so that Eq.\ \autoref{eq:mass} gives proper mass values for the star. $M(R)$ relations are shown in Fig.\ \autoref{fig:mr_zero} for the three different EOS considered in this study. At this point it seems that it is not possible to put constraints on the coupling constant and the central value for the scalar field separately, but the combination $\xi \phi_{\rm c}^2$ can be restricted. Parameter spaces of the combination based on the results obtained in Fig.\ \autoref{fig:mr_zero} are illustrated in Fig.\ \autoref{fig:constraints}. 

As a result of our numerical analysis, we have found that it is possible to write $\phi_\infty \!\approx\! \phi_{\rm c}$ since the deviation between the central and the asymptotic values of the scalar field are in the same order of magnitude. Another point that deserves to be mentioned for this model is that for all EOS with a fixed value of the coupling constant, the radius of the star is slightly altered depending on the central value of the scalar field, while the maximum mass has no significant change.

Finally, we compare our findings on the term $\xi\phi_{\rm c}^2$ with the constraint for $\xi$ given in Ref.\ \cite{hrycyna_what_xi} and the results are summarized in Table \autoref{tab:coup_constr_cosmo}. We see that the allowed range for the central value of the scalar field must be relatively small and that the $M(R)$ curves for the positive values of the coupling constant have to be discarded. Although we give only positive ranges for $\phi_{\rm c}$ in the table, the constraints are mathematically valid for negative values as well. For WFF1 the range represents only the limits, since there are values that are not applicable to $\phi_{\rm c}$ in this range.

\vspace{5mm}

\def\arraystretch{1.1}
\begin{table}[!t]
	\centering
	\caption{Constraints on the central value of the scalar field for SLy, WFF1, and MS1. The results are obtained by combining the numerical values of the term $\xi\phi_{\rm c}^2$ given in Fig.\ \autoref{fig:constraints} and the restriction for $\xi$ found in Ref.\ \cite{hrycyna_what_xi}. An approximation $\phi_{\rm c} = \phi_\infty$ is made for these constraints as explained in the text.} \vspace{2mm}
	\label{tab:coup_constr_cosmo} 
	\begin{tabular}{c|c|c|c|}
		\cline{2-4}
		& $\qquad\qquad$\textbf{SLy}$\qquad\qquad$ & $\qquad\qquad$\textbf{WFF1}$\qquad\qquad$ & $\qquad\qquad$\textbf{MS1}$\qquad\qquad$ \\ \hline
		\multicolumn{1}{|c|}{$\qquad \xi\phi_{\rm c}^2 \qquad$} & (-0.01, 0) & (-0.0155, -0.003) & (0.008, 0.027) \\ \hline\hline
		\multicolumn{1}{|c|}{$\xi$} & \multicolumn{3}{c|}{(-2.6051, -0.0633)} \\ \hline\hline 
		\multicolumn{1}{|c|}{$\qquad \phi_{\rm c} \qquad$} & (0, 0.398) & (0.034, 0.495) & - \\ \hline
	\end{tabular} \vspace{3mm}
\end{table}

\subsection{Higgs-like Potential}
Here we examine the following Higgs-like symmetry breaking potential
\begin{equation}
	V(\phi) = \sfrac{\lambda}{4} \big(\phi^2 - \nu^2 \big)^2
\label{eq:higgs_pot}
\end{equation}
that was first examined in Ref.\ \cite{fuzfa1}, where a constant density configuration was considered and values of the potential parameters have been taken as $\lambda \!\sim\! 0.1$ and $\nu \!=\! 246$ GeV. In our study we use realistic EOS and we implement the same numerical technique for matching the solutions at the surface of the star as in the previous case and observe whether this approach gives reasonable $M(R)$ relations together with asymptotic flatness. Additionally, instead of the vacuum expectation value for the Higgs field which is $\nu \!\approx\! 10^{-55}$ km in geometrical units, we use the values that are comparable with the zero potential case. Besides the convenience for numerical calculations, this choice will allow us to use the constraints obtained in the previous case and to see the relation between the coupling constant and the parameter $\nu$.

\begin{figure*}[t]
	\centering
	\begin{tabular}{@{}c@{}}
		\includegraphics[width=.5\linewidth]{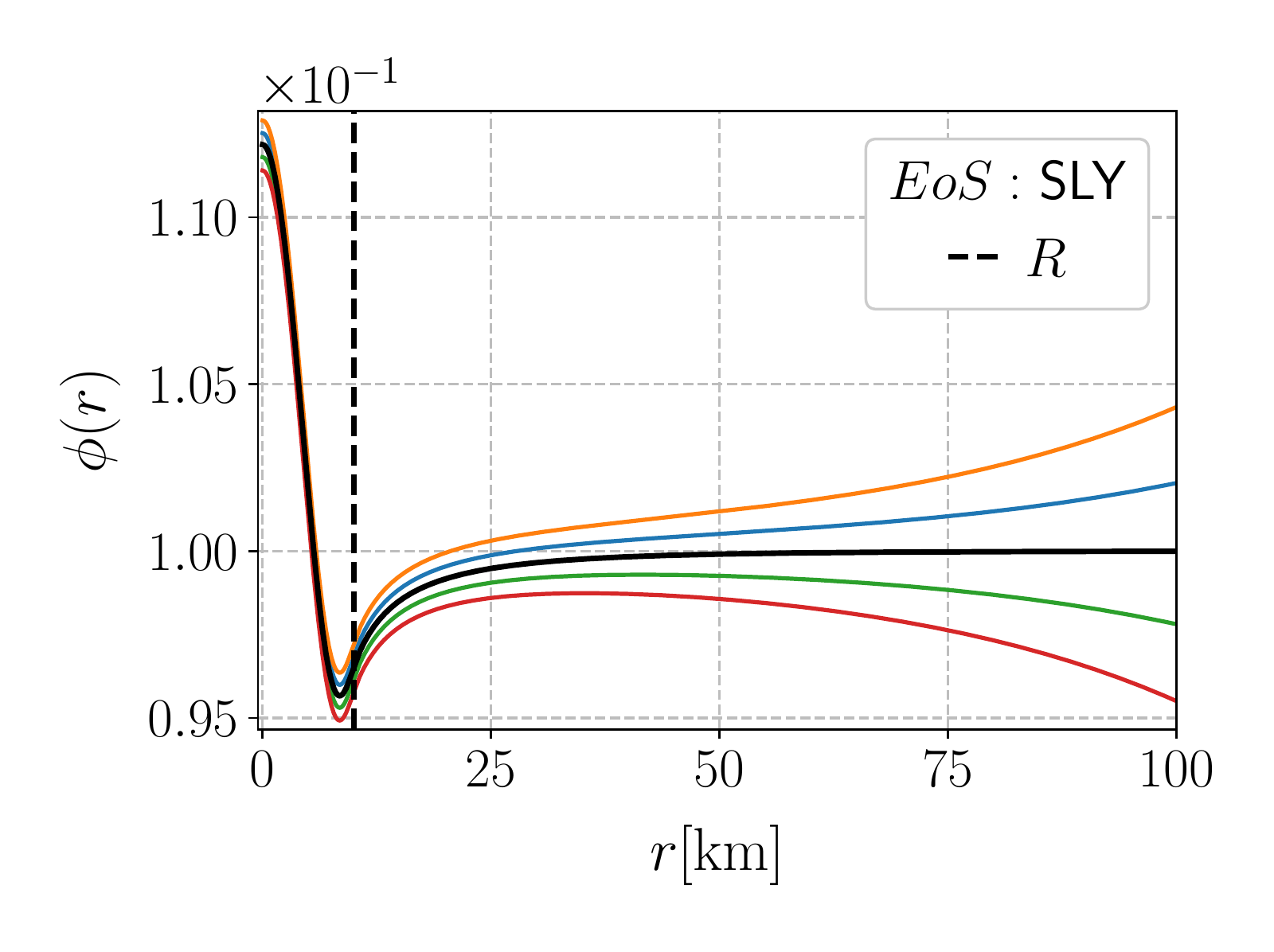} \hspace{-5mm} \\
	\end{tabular}
	\begin{tabular}{@{}c@{}}
		\includegraphics[width=.5\linewidth]{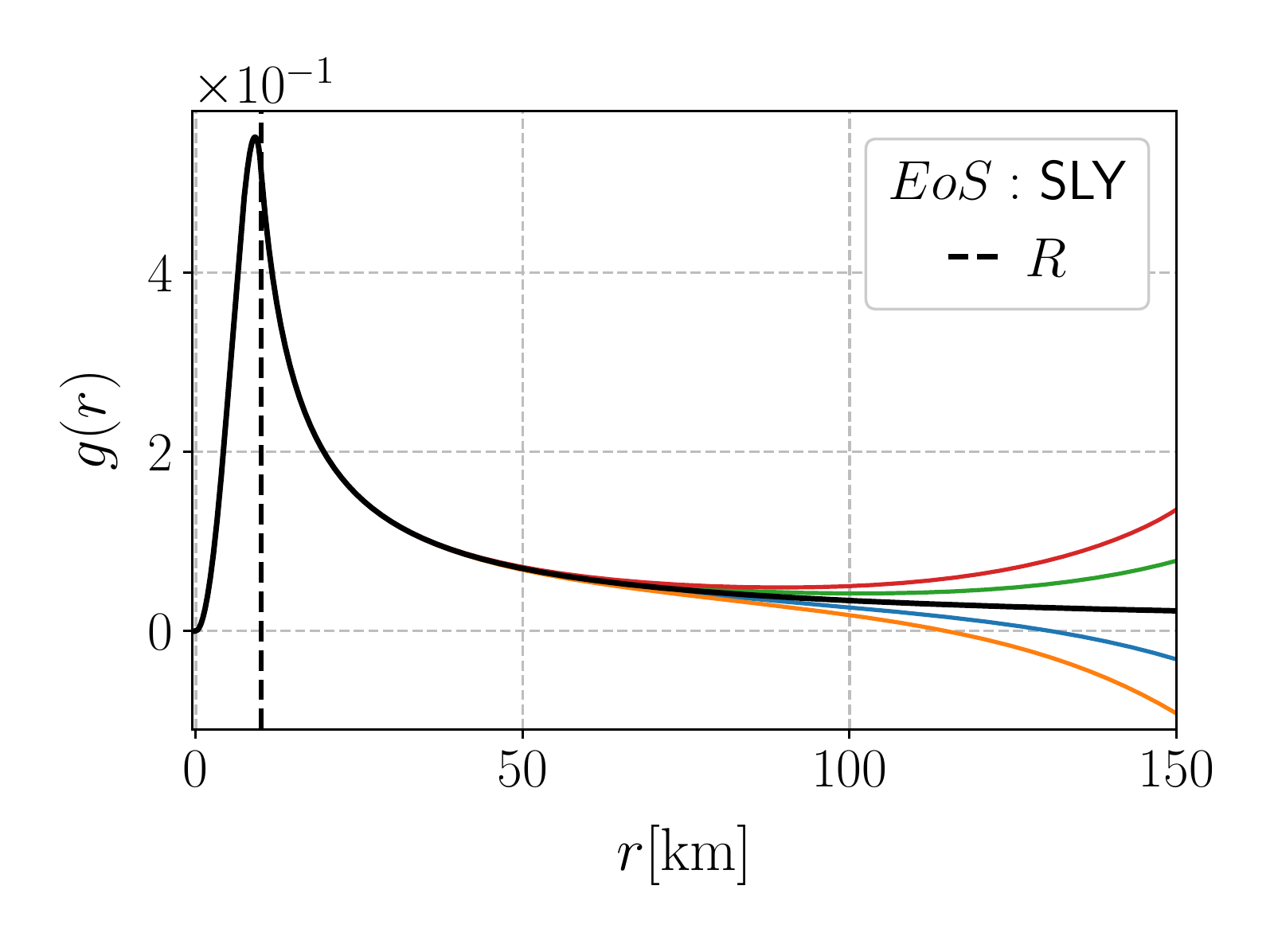} \\
	\end{tabular}
	
	\vspace{-4mm}
	
	\caption{Radial profiles of the scalar field $\phi(r)$ (left) and the metric function $g(r)$ (right) for Higgs-like potential given in Eq.\ \autoref{eq:higgs_pot}. Here $\xi=-0.65$ and $\lambda \!=\! \nu \!=\! 0.1$ and $R$ indicates the radius of the star. The solutions with different colors are obtained by different central values of the scalar field and the color coding is consistent for two figures. Only the solution shown as the black curve has the correct asymptotic behavior.}
	\label{fig:higgs_func} \vspace{2mm}
\end{figure*}

Based on the analogy explained in Ref.\ \cite{chameleon_cosmo}, we can write Eq.\ \autoref{eq:fieldeq2} in the form
\begin{equation}
	\boxempty \! \phi = - \dfrac{d V_{\rm eff}}{d \phi},
\end{equation}
where we define the effective potential as
\begin{equation}
	V_{\rm eff} = \sfrac{1}{2} \xi \R \phi^2 - \sfrac{\lambda}{4} \big(\phi^2 - \nu^2 \big)^2 \: .
\end{equation}
Inside the star $\R \!\neq\! 0$ so that the critical points are $\phi_{\rm cr} \!=\! 0, \pm \sqrt{\nu^2 + \xi \R / \lambda}$. On the surface of the star ($\R \!=\! 0$) they become $\phi_{\rm cr} \!=\! 0,\pm \nu$, the first of which is stable whereas the others are unstable critical points of the effective potential (for details see Ref.\ \cite{fuzfa2}). Although $\phi_{\rm cr} \!=\! 0$ is the stable point of the potential, it does not satisfy $V(\phi_{\rm cr}) \!=\! 0$, which is the first necessary condition for the stabilization of the mass integral given in Eq.\ \autoref{eq:mass} and in this case it is not possible to obtain a stable star configuration. Thus we consider the unstable critical points as in Refs. \cite{fuzfa1,fuzfa2} and this, in turn, forces one to choose a unique initial condition for the scalar field at the center, such that the asymptotic values of the metric functions and the scalar field satisfy the asymptotic flatness condition. As an example, we consider the SLy in Fig.\ \autoref{fig:higgs_func} for $\xi \!=\! -0.65$. It seems that the valid profile of the metric function $g(r)$ and the scalar field, which are shown as black curves on the figure, are quite similar to those of the zero potential case given in Fig.\ \autoref{fig:zero_func}. On the right panel of the same figure, solutions above the black curve are divergent, whereas others roll down to the stable critical point ($\phi_{\rm cr} \!=\! 0$). This result is compatible with Refs.\ \cite{fuzfa1,fuzfa2} where the solution curves inside the star are different than ours because only constant density was considered there.

We find small deviations on $M(R)$ relations of the star as well as on the radial profile of the scalar field and the metric functions compared to the zero potential case which is in accordance with Refs.\ \cite{fuzfa1,fuzfa2}. A sample of numerical results is given in Table \autoref{tab:pot_compare} for different values of the parameters: we either keep the parameter $\nu$ constant and vary the order of magnitude of the $\lambda$, or vice versa.  In the case where $\lambda$ is kept constant we have considered the values of $\nu$ that do not contradict the restriction $\kappa_{\rm eff}>0$, which implies $1 + \kappa \xi \phi^2 > 0$. This condition for the scalar field $\phi$ can be transformed to that of the parameter $\nu$ since the radial variation of the scalar field does not deviate significantly from the critical value of the potential and we can write $1 + \kappa \xi \nu^2 \gtrsim 0$.  If the numerical values ($\kappa \!=\! 8\pi$, $\xi \!=\!-0.65$) are used in this relation, it becomes $\nu \lesssim 0.25$, which explains the chosen values of $\nu$ in the table.

\renewcommand{\arraystretch}{0.9}
\begin{table}[t]
	\centering
	\caption{Mass and radius values of the configuration for two different potentials with various $\nu$ and $\lambda$ values for the Higgs-like potential. Here $\xi$ is set to $-0.65$ for SLy EOS. The same initial conditions for the central pressure are used for all cases. Resulting numerical values of mass, radius and central value of the scalar field are rounded up to an order that still shows the differences.}
	\label{tab:pot_compare} \vspace{3mm}
	\begin{tabular}{|c||c|c|c|c||c|c|} \hline
		& \multicolumn{4}{|c||}{\textbf{Higgs-like Potential}}   & \multicolumn{2}{c|}{\textbf{Zero Potential}} \\ \hline \hline
		$\qquad \phi_{\rm c} \qquad$ & $\qquad \nu \qquad$ & $\qquad \lambda \qquad$ & $\quad M\rm[M_\odot] \quad$ & $\quad R$[km] $\quad$   &  $\quad M\rm[M_\odot] \quad$ & $\quad R$[km] $\quad$ \\ \hline \hline
		0.1122 & \multirow{4}{*}{0.1} & 0.1 & 2.22 & 10.01 & 2.18 & 10.01 \\ 
		0.1121 & & 1 & 2.21 & 10.04 & 2.18 & 10.01 \\ 
		0.1083 & & 10 & 2.20 & 10.08 & 2.17 & 9.94 \\ 
		0.1020 & & 100 & 2.19 & 10.12 & 2.15 & 9.85 \\ \hline \hline
		0.0579 & 0.05 & \multirow{4}{*}{0.1} & 2.05 & 9.41 & 2.04 & 9.40 \\  
		0.1122 & 0.1 &             & 2.22 & 10.01 & 2.18 & 10.01 \\ 
		0.1658 & 0.15 &             & 2.59 & 11.54 & 2.54 & 11.46 \\ 
		0.2242 & 0.2 &             & 3.82 & 15.87 & 3.75 & 15.58 \\ \hline
	\end{tabular} \vspace{2mm}
\end{table}

Finally, we compare the results for the potentials considered. We first notice from Table \autoref{tab:pot_compare} that the mass of the star increases with larger central values of the scalar field for both potentials and all EOS. This is valid for the radius as well, except for the Higgs-like potential with fixed $\nu$ for which the radius gets smaller with decreasing $\lambda$ values. Slightly more massive configurations are obtained for the Higgs-like potential for the same central value of the scalar field; that is, the presence of the scalar field potential makes the star more massive. Besides that we see no significant effect of the parameter $\nu$ on the solutions. On the other hand, the parameter $\lambda$ has an intriguing impact on the configuration: it makes the relation between the mass and the radius inversely proportional unlike other cases. Additionally, the asymptotic value of the scalar field must be equal to the value of $\nu$ as explained before and indeed Table \autoref{tab:pot_compare} shows that there is no significant difference between the central value of the scalar field and the value of $\nu$ chosen. Thus, we can write $\phi_{\rm c} \approx \nu$ to get
\begin{equation}
	- 0.01 < \xi\nu^2 < 0 \: , \quad - 0.0155 < \xi\nu^2 < -0.003 \quad \mbox{and} \quad 0.008 < \xi\nu^2 < 0.027 \:
\label{eq:higgs_constraints}
\end{equation}
for SLy, WFF1 and MS1, respectively. In particular, for the case shown in Fig.\ \autoref{fig:higgs_func}, where $\nu \!=\! 0.1$, we get $-1 < \xi < 0$, $-1.55 < \xi < -0.3$, and $0.8 < \xi < 2.7$. On the other hand, combining Eq.\ \autoref{eq:higgs_constraints} with the cosmological restriction \cite{hrycyna_what_xi}, as in the previous section, makes the constraints for $\phi_{\rm c}$ given in Table \autoref{tab:coup_constr_cosmo} also viable for $\nu$.

\section{CONCLUSION}  \label{conclusion}
In this study we have investigated the neutron star structure in the presence of a nonminimally coupled scalar field. The static and spherically symmetric solutions have been examined for three realistic EOS and the two cases, the zero potential and the Higgs-like potential which allows one to construct stable stellar solutions with the choice of a unique central value for the scalar field. The zero potential case has been subjected to many studies in the context of spontaneous scalarization and, in general, the Einstein frame has been used with some specific coupling terms (e.g.\ Refs.\ \cite{Damour:1992,damour1993,damour_ppn,farese2004,asymmetron} and references therein). However, we have made the analysis and numerical calculations in the physical Jordan frame and, in this manner, we have intended to discuss possible constraints on the parameters directly without referring to the equivalence of the two frames.

The model with zero potential was studied in Ref.\ \cite{kazanas2014} only for the conformal coupling in isotropic coordinates and the results were compared with GR solutions.  Spontaneous scalarization encountered in this model was investigated in Ref.\ \cite{salgado1998} where only asymptotically vanishing values of the scalar field were considered. Besides the cases where contribution of the potential is neglected, the Higgs-like potential has been studied for the same model in Refs.\ \cite{fuzfa1,fuzfa2} where only a constant density configuration was taken into account and in order to examine the possibility of the existence of solutions, some assumptions and approximations were adopted. However, in our study, we have not made any assumptions on the solutions and used realistic EOS to obtain $M(R)$ relations of neutron stars. We have implemented a numerical approach for matching the interior and the exterior solutions of the star and put constraints on the term $\xi\phi_{\rm c}^2$ from the observational limits on the mass and the radius of neutron stars. The restrictions on the central value of the scalar field and the value of the coupling constant cannot be determined separately in this approach. Thus to put a limit on $\phi_{\rm c}$, we have considered the results of Ref.\ \cite{hrycyna_what_xi}, where the coupling constant $\xi$ was restricted with a cosmological point of view. We have also reviewed the PPN formalism and dipole radiation in pulsar-white-dwarf binary systems and discussed the emerging constraints briefly.

We also discuss the differences of the zero and the Higgs-like potentials and compare the results by comparing the mass and the radius of the same configurations, which are summarized in Table \autoref{tab:pot_compare}. The presence of a scalar potential in the model increases the stellar mass. The characteristic radial profile of the scalar field is similar in the two cases, as can be seen by comparing Figs.\ \autoref{fig:zero_func} and \autoref{fig:higgs_func}. The parameters of the Higgs-like potential have been analyzed as well. As shown in Table \autoref{tab:pot_compare} there is no unexpected effect due to the parameter $\nu$, which determines the asymptotic value of the scalar field. The other parameter $\lambda$, on the other hand, has an evident effect on the radius of the star. Although the stellar mass increases with the central value of the scalar field as in the case of zero potential, the radius of the star increases with $\lambda$. We also notice that the constraints obtained for zero potential for $\phi_{\rm c}$ can be applied to the case with the Higgs-like potential, since the behavior of the scalar field is not too much different. Thus the parameter $\nu$ appearing in the potential can be substituted for the central value of the scalar field, so that the restrictions given in Eq.\ \autoref{eq:higgs_constraints} are obtained.

The restrictions for the parameters in our model are obtained by using the observational mass value of \textit{J0348+0432}, that is $(2.01\pm0.04)\,\rm M_{\odot}$ \cite{J0348}, and the radius constraint coming from \textit{GW170817} \cite{LIGO_GW170817}. Combining our analysis with other methods, such as the one given in Ref.\ \cite{hrycyna_what_xi} in the cosmological context, could allow one to infer each parameter of the model separately rather than the particular combinations given here. The improvement of the data with the future more precise observations may restrict the parameters of the model further.

\bibliographystyle{apsrev4-1}
\bibliography{references}

\end{document}